\documentclass{aastex631}

\usepackage{nicefrac}
\usepackage{amsmath}

\hypersetup{linkcolor=red,citecolor=blue,filecolor=cyan,urlcolor=teal}

\submitjournal{PSJ}
\accepted{June 3rd, 2022}

\shorttitle{Atmospheric Tides on M~Earths}
\shortauthors{Navarro et al.}

\graphicspath{{./}{figures/}}

\turnoffeditone

\begin{document}

\title{Atmospheric gravitational tides of Earth-like planets orbiting low-mass stars}

\correspondingauthor{Thomas Navarro}
\email{thomas.navarro@mail.mcgill.ca}
\author[0000-0002-0218-6617]{Thomas Navarro}
\affiliation{McGill Space Institute, Montr\'eal, Canada}
\affiliation{Dept of Earth, and Planetary Sciences, McGill University, Montr\'eal, Canada}

\author[0000-0002-5593-8210]{Timothy M.\ Merlis}
\affiliation{Program in Atmospheric and Oceanic Sciences, Princeton University, Princeton, USA}

\author[0000-0001-6129-5699]{Nicolas B.\ Cowan}
\affiliation{McGill Space Institute, Montr\'eal, Canada}
\affiliation{Dept of Earth, and Planetary Sciences, McGill University, Montr\'eal, Canada}

\author[0000-0002-2463-7917]{Natalya Gomez}
\affiliation{McGill Space Institute, Montr\'eal, Canada}
\affiliation{Dept of Earth and Planetary Sciences, McGill University, Montr\'eal, Canada}

\begin{abstract}

    Temperate terrestrial planets orbiting low-mass stars are subject to strong tidal forces. The effects of gravitational tides on the solid planet and that of atmospheric thermal tides have been studied, but the direct impact of gravitational tides on the atmosphere itself has so far been ignored. We first develop a simplified analytic theory of tides acting on the atmosphere of a planet.  We then implement gravitational tides into a general circulation model of a \edit1{static-ocean} planet in a short-period orbit around a low-mass star---the results agree with our analytic theory. Because atmospheric tides and solid-body tides share a scaling with the semi-major axis, we show that there is a maximum amplitude of the atmospheric tide that a terrestrial planet can experience while still having a solid surface; Proxima Centauri b is the poster child for a planet that could be geophysically Earth-like but with atmospheric tides more than 500$\times$ stronger than Earth's. In this most extreme scenario, we show that atmospheric tides significantly impact the planet's meteorology---but not its climate. \edit1{Two possible modest climate impacts are enhanced longitudinal heat transport and cooling of the lowest atmospheric layers. The strong radiative forcing of such planets dominates over gravitational tides, unlike moons of cold giant planets, such as Titan.} We speculate that atmospheric tides could be climatologically important on planets where the altitude of maximal tidal forcing coincides with the altitude of cloud formation and that the effect could be detectable for non-Earth-like planets subject to even greater tides.

\end{abstract}

\keywords{exoplanet atmospheres --- tides -- M~Earth}

\section{Introduction}

\edit1{Low-mass} stars are prime targets for understanding and characterizing temperate terrestrial exoplanets. M dwarfs represent 75\% of stars because they form more often and live much longer than more massive stars \citep{bochanski2010,reid2005}.  Moreover, temperate terrestrial planets are more common around M dwarfs than Sun-like stars \citep{dressing2015} and  
these planets are \edit1{potentially} easier to characterize than Earth-like planets orbiting Sun-like stars \citep{charbonneau2017}. Establishing whether temperate rocky planets orbiting M dwarfs, the so-called M~Earths, are actually habitable is the most pressing question in the search for extrasolar life \citep{shields2016}.

M~Earths are subject to much stronger gravitational tides than the Earth. 
Although M dwarfs are less massive than the Sun, their much lower luminosity puts their habitable zone at semi-major axes of 0.01--0.1~AU \citep{kopparapu2013}, resulting in tidal forces many orders of magnitude greater than those on Earth.  
As a result, planets orbiting \edit1{low-mass stars} are most likely tidally locked, meaning that there is a resonance between the planet's spin rate and its orbital rotation rate \citep{dole1964}. The prospect of tidal locking into a synchronous rotation---with permanent day and night hemispheres---was originally thought to be an impediment to M~Earth habitability \citep{kasting1993}. 
The possible climates of M~Earths have subsequently been explored with global climate models (GCMs) and results suggest that they could remain habitable \citep{joshi1997,Shie:19}, whether they rotate synchronously \citep{Merl:10,2013ApJ...771L..45Y,Turb:16,Pier:19} or not \citep{Turb:16}.

The effects of gravitational tides on the atmospheric circulation of M~Earths has so far been ignored in exoplanetary studies. 

\citet{Barn:13} discussed the role of tidal forces on the solid planet, finding that tidal heating may be comparable to instellation for planets with nonzero orbital eccentricity and hence could trigger 
a runaway greenhouse, especially for planets orbiting in the habitable zones of stars less massive than 0.3 $M_\odot$.
\citet{ChapLindz} developed a theory for the gravitational tides of the Moon on Earth's atmosphere.
Generalizing this theory to other planets, an ab~initio analytical model was developed by \citet{Aucl:17}, mostly focused on atmospheric torque and the thermal tides, but while their theory included gravitational effects, it did not address the effect of gravitational tides on atmospheric variables.
Moreover, their model cannot account for atmospheric circulation and important meteorological phenomena such as clouds.

The only attempts to simulate the effects of gravitational tides on a planetary body's atmospheric circulation have been for Saturn's moon Titan \citep{Toka:02,Char:22}.
These are the only examples of the inclusion of tidal forces in an atmospheric GCM.
\citet{Toka:02} found a substantial effect on the circulation, decreasing the superrotating winds at 100 km altitude from 25 m~s$^{-1}$ to 8 m~s$^{-1}$.
\citet{Walt:06} explored the transport of aerosols and the formation of a haze by Saturnian tides on Titan.
\edit1{\citet{Char:22} found very weak tidal winds in the lower troposphere of Titan, but with a noticeable fluctuation of up to 4~Pa of surface pressure, enough to constrain the deformation of Titan if pressure were measured in-situ}.

In this paper, we aim to address the effects of gravitational tides on the meteorology and climate of synchronously rotating M~Earths by means of numerical modeling.
Section~\ref{sec:potential} presents a simple analytical relation between tidal potential and surface pressure, section~\ref{sec:umax} explores the upper bound of atmospheric tides on M~Earths, while section~\ref{sec:others} discusses other tidal configurations not further explored in this article. Section~\ref{sec:simus} presents GCM simulations including gravitational tidal forces, section~\ref{sec:results} details their effects on the circulation and climate, and we conclude in section~\ref{sec:conclusion}. \\


\section{Gravitational potential and surface pressure} \label{sec:potential}

\begin{figure}
\centering
\includegraphics[height=0.9\textheight]{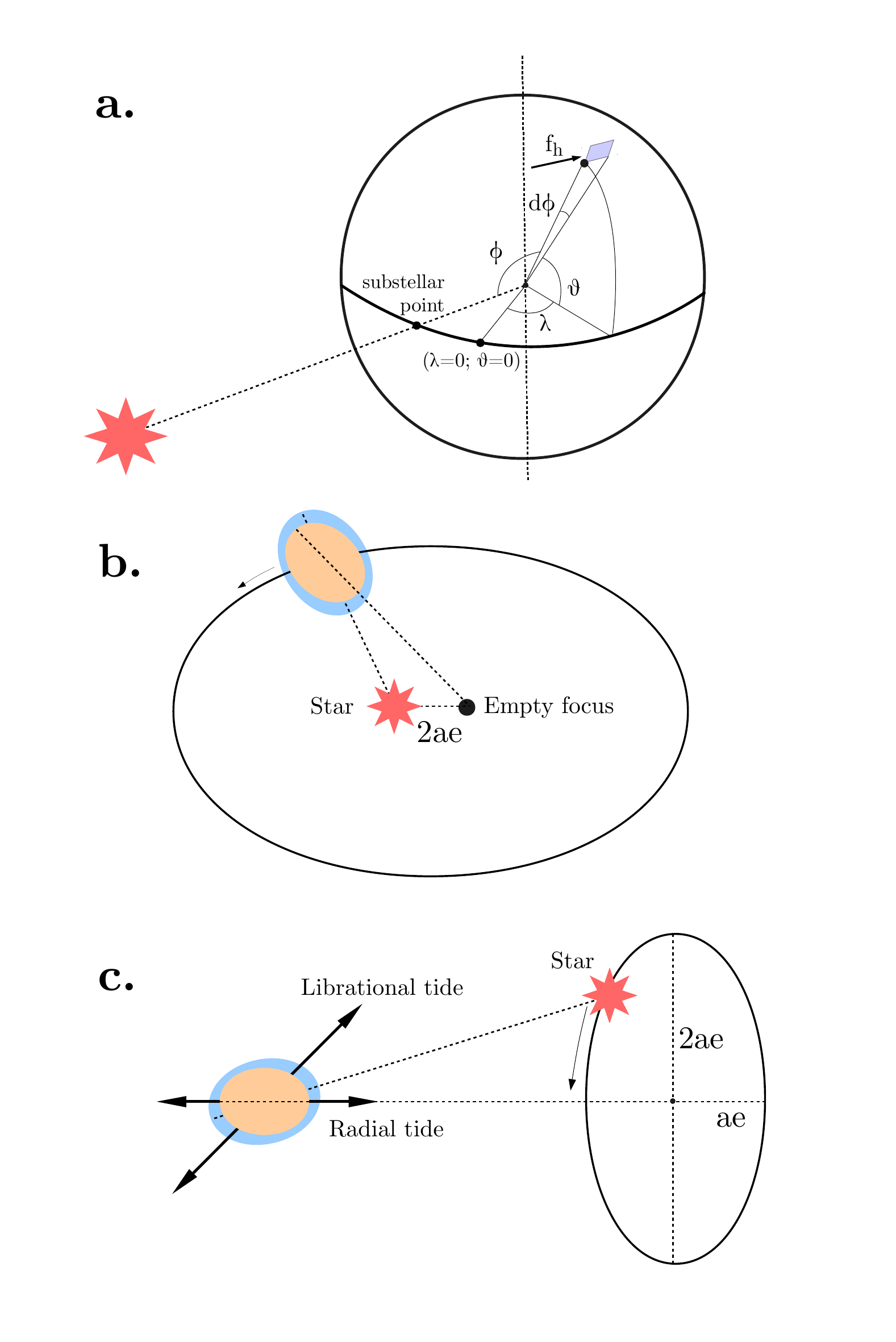}
\caption{Schematics of atmospheric tides. (a) Planetary and tidal coordinates. (b) Tides in a star-centered inertial frame. (c) Tides in a planet-centered frame. In the limit of small eccentricity ($e\ll1$), the planet's constantly deformed solid body points to the orbit's empty focus, generating a traveling tidal wave for a 1:1 spin-orbit resonance (a.k.a.\ synchronous rotation). Adapted from \cite{Murr:99}, \citet{Lore:92}, and \citet{Toka:02}.}
\label{fig:schematics}
\end{figure}

Let $U$ be the tidal potential at a point on a planet.
We assume a shallow atmosphere, such that all altitudes in the atmosphere at some location are the same distance $R_p$ from the planet's center.
This assumption is usually valid for terrestrial atmospheres, while the depth of the atmosphere should be taken into account in the case of giant planets or Saturn's moon Titan \citep{Tort:14}.
Thus, the tidal force per unit mass has a horizontal component at any given point in the atmosphere:
\begin{linenomath*}
\begin{equation}
 f_h = \frac{1}{R_p} \frac{\partial U}{\partial \phi},
\end{equation}
\end{linenomath*}
where $R_p$ the planetary radius and $\phi$ is the angle formed by the position vectors of the point and the star, with the center of the planet as the origin (i.e., $\phi$ is the stellar zenith angle at that location on the planet). Note that $\cos\phi=\cos\theta\cos\lambda$, where $\theta$ and $\lambda$ are the latitude and longitude of the point, with $\lambda = 0$ at the substellar meridian~(Figure \ref{fig:schematics}).

As in \citet{Lore:92}, we write the force balance of a fluid parcel:
\begin{linenomath*}
\begin{equation}
 dp = f_hR_p\rho d\phi,
\end{equation}
\end{linenomath*}
with $p$ the pressure and $\rho$ the density.
Using the ideal gas law, 
\( \displaystyle \rho=\frac{pm}{RT} \), 
with $T$ the temperature, $m$ the molecular mass, and $R$ the ideal gas constant, we get \edit1{the equilibrium tide \citep[e.g.,][]{Ogil:04,Ogil:14}}:
\begin{linenomath*}
\begin{equation}
\label{eq:dpsurp}
 \frac{dp}{p} = \frac{m}{RT} dU.
\end{equation}
\end{linenomath*}
\edit1{Dynamical tides obtained from linear tidal theory \citep{Aucl:17} are not considered in this simple analytical model.}
For a rigid planet, the Legendre polynomial expansion of the tidal potential \( \displaystyle U = -\frac{GM_*}{d}\) is \citep[e.g.,][]{TideBook}
\begin{linenomath*}
\begin{equation}\label{eq:totpotential}
 U = -\frac{GM_*}{d} \left[ \frac{3\cos^2\phi-1}{2} \left({\frac{R_p}{d}}\right)^2 + \mathcal{O}\left(\frac{R_p}{d}\right)^3 \right],
\end{equation}
\end{linenomath*}
with $G$ the gravitational constant, $M_*$ the star mass, and $d$ the star--planet distance.
Integrating equation \eqref{eq:dpsurp} from the substellar point $\phi'=0$ to any location $\phi'=\phi$, ignoring higher-order terms, and assuming an isothermal atmosphere, the surface pressure field $p_s$ can be written to leading order as
\begin{linenomath*}
\begin{equation}
\ln\left(\frac{p_s(\phi)}{p_{\rm high}}\right) = -\kappa \sin^2\phi,
 \label{eq:psampl}
\end{equation}
\end{linenomath*}
with $p_{\rm high}=p_s(\phi=0)$ the high-tide surface pressure at the substellar point, and the amplitude of the atmospheric tide is \( \displaystyle \kappa = \frac{3}{2} \frac{m}{RT} \frac{GM_*R_p^2}{d^3} \).\\

Tidal forces act on both the solid planet and its atmosphere, so we need to consider the difference between the tidal forces in the atmosphere and the solid body. A static tidal force field \edit1{does not directly induce atmospheric motions, although the atmospheric flow (caused by, e.g., radiative forcing) may be indirectly impacted by tidal potential variations.} 
Indeed, there are no \edit1{time-varying} tidal forces in the atmosphere of a synchronously rotating planet on a circular orbit.
A time-variable tide, on the other hand, distorts the atmosphere faster than the solid body, resulting in a time-dependent component of the net atmospheric tidal force. 
Time-varying tidal forces may come from the synchronous rotation of an eccentric planet \edit1{\citep{Murr:99}}, asynchronous rotation (whether or not the orbit is circular; e.g., \edit1{\citet{Leco:15}}), or tidal forces of another planet \edit1{\citep{Hami:19}}.  In the current study, we limit ourselves to the first case.

A synchronously rotating planet always shows the same side to its star.
If the planet's orbit is circular, it has zero obliquity and the planet has relaxed to its prolate shape (i.e., the solid body is deformed), then the gravitational atmospheric tide 
\edit1{in the solid body and atmosphere are subject to the same static forces. Thus, the solid-body deformation cancels out the equilibrium tide in the atmosphere.}

However, if the orbit is eccentric, the distance to the star varies\edit1{, which raises a radial tide}, and the substellar point moves longitudinally\edit1{, which raises a librational tide}, thus creating a nonzero time-dependent component \edit1{\citep{Murr:99}};
this is the case with Titan, with Saturn in the role of the attracting body \edit1{\citep{Toka:02}}.

If a planet has a small eccentricity, zero obliquity, and has relaxed to its prolate shape, then there is a \edit1{time-varying} gravitational tide because the planet keeps one side pointed to the empty focus, rather than the host star, as it orbits \citep[][and Figure \ref{fig:schematics}]{Murr:99}.
In that case, \citet{Saga:82} and \citet{Murr:99} give us the formulation of the time-dependent part of the potential for small eccentricities (e$\ll$1):
\begin{linenomath*}
\begin{equation}
\label{eq:tipotential}
 U = -\frac{GM_*R_p^2}{a^3} 3 e \left[\frac{3 \cos^2\theta \cos^2\lambda -1}{2} \cos(nt) + \cos^2\theta\cos^2(2\lambda)\sin (nt) \right]
\end{equation}
\end{linenomath*}
with \edit1{$t$ the time in units of orbital period,} $\theta$ the latitude, $\lambda$ the longitude taken from the mean substellar point, $a$ the semi-major axis, $e$ the eccentricity, and $n$ \edit1{the mean motion.}
If the time-dependent Equation~\ref{eq:tipotential} is used rather than 
 Equation~\ref{eq:totpotential}, the amplitude of the atmospheric tide from Equation~\ref{eq:psampl} can be written as a function of orbital parameters $a$ and $e$:
\begin{linenomath*}
\begin{equation}
\label{eq:ampl2}
\kappa = \frac{9}{2}\frac{m}{RT} \frac{GM_*R_p^2}{a^3} e.
\end{equation}
\end{linenomath*}

In order to get a sense of how the tide travels across the planet, it can be useful to move to a planet frame~(Figure~\ref{fig:schematics}), where the star moves along an ellipse of dimension ($ae$ ; $2ae$).
The star gives rise to the radial tide (first term in Equation~\ref{eq:tipotential}) and the librational tide (due to the second term in Equation~\ref{eq:tipotential}).
As for the Earth-Moon tide, the $ \cos(nt)$ shows that the tide of a planet in synchronous rotation has a longitudinal wavenumber 2--two highs and two lows at any given time.
However, and unlike Earth-Moon tides, from the $\cos^2\lambda$ term, there is one peak and low per period at any location \citep[for examples on Titan, see][]{Toka:02}.

\edit1{Equation \ref{eq:tipotential} is the potential exerted on the whole planet. The tidal potential exerted on the atmosphere alone is actually reduced and phase-shifted by the deformation of the solid body. On Titan, a moon with a surface ice shell and subsurface ocean, the potential is reduced to less than 20\% of the tidal potential of Equation~\ref{eq:tipotential}~\citep{Char:22}. In our case, we do not take into account this reduction of the tidal potential, as we explore the maximal impact of tides, as detailed below.}

\section{Interior tidal heating and maximal surface tidal potential} \label{sec:umax}

We limit our study to planets with solid surfaces. Interior heating by tidal dissipation in the solid planet is a limiting factor, lest the surface melt, and gives a maximum tidal potential.  
\edit1{A planet with a molten surface would have orders of magnitude more surface deformation. Moreover, it would be unlikely to be habitable.}

Let us consider two constraints to find the maximum value of tidal potential amplitude: a maximum interior tidal heating and a maximum eccentricity.
From \citet{Dris:15} and \citet{Sega:88}, the surface heat flux $q$ from interior tidal heating is
\begin{linenomath*}
\begin{equation}
q = \frac{21}{8\pi} \text{Im}(k_2) G^{\nicefrac{3}{2}} M_*^{\nicefrac{5}{2}} R_p^3 e^2 a^{-\nicefrac{15}{2}}
 \label{eq:qtidal}
\end{equation}
\end{linenomath*}
with $k_2$ the tidal Love number.

From Equation \ref{eq:qtidal}, we isolate the quantity $M_*a^{-3}$ and obtain the maximum tidal potential amplitude using Equation~\ref{eq:tipotential}:
\begin{linenomath*}
\begin{equation}
U_{\rm max}  = 3 { \left(\frac{8\pi q_{\rm max}}{21 \text{Im}(k_2)}\right) }^{\nicefrac{2}{5}}  G^{\nicefrac{2}{5}} R_p^{\nicefrac{4}{5}} e_{max}^{\nicefrac{1}{5}}
 \label{eq:Umax}
\end{equation}
\end{linenomath*}
If we rewrite Equation~\ref{eq:psampl} for a deformable planet in synchronous rotation, we obtain the amplitude of surface pressure change of an isothermal atmosphere in synchronous rotation:
\begin{linenomath*}
\begin{equation}
\ln\left(\frac{p_{\rm low}}{p_{\rm high}}\right) = -\frac{m}{RT} U_{\rm max}. \label{eq:psamplsynchro}
\end{equation}
\end{linenomath*}
Equation \ref{eq:Umax} shows that the maximal surface tidal potential depends mostly on the planetary radius and is less sensitive to assumed values of $e_{\rm max}$ or $q_{\rm max}$.
\edit1{Note that these surface pressure changes ignore the impact of solid-body deformation on the potential exerted on the atmosphere \citep[see, e.g.,][for Titan]{Char:22}.}

\begin{figure}[ht]
\centering
\includegraphics[width=0.9\textwidth]{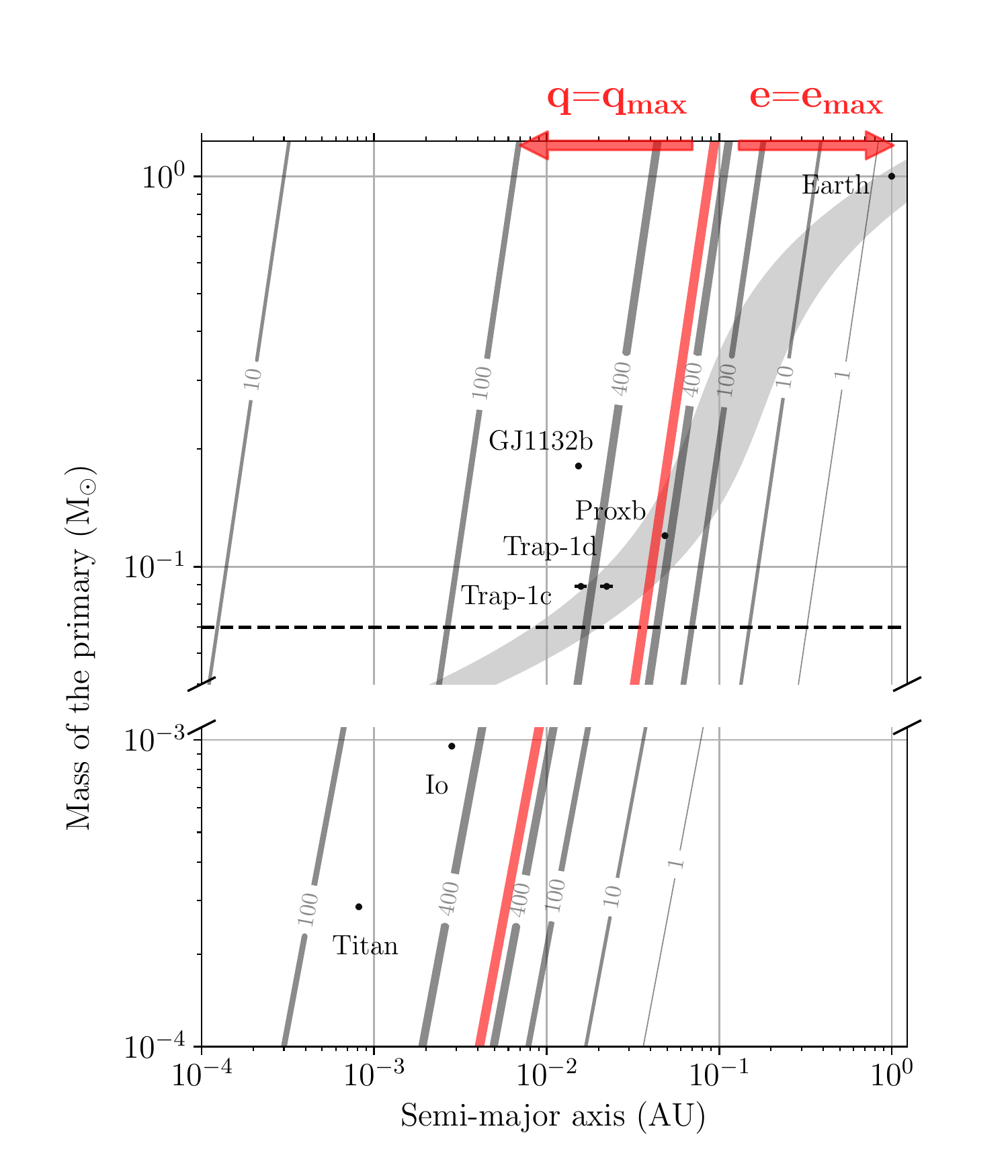}
\caption{Example of tides in the mass-distance space. Maximal tidal potential at the surface of an Earth-sized planetary body is shown (gray lines, in N~m~kg$^{-1}$) with a maximal eccentricity $e_{\rm max}=0.1$ and with the instellation habitable zone shown with gray shading \citep{kopparapu2013}.
The maximal tidal potential $U_{\rm max}$=707 N~m~kg$^{-1}$ is shown with a red line.
For a smaller semi-major axis, stronger tides produce an interior heat flux (greater than Io's) no longer negligible for the planetary climate.
\edit1{Thus, for smaller semi-major axis and higher primary mass than the ones of $U_{\rm max}$, the eccentricity is given by Equation \ref{eq:qtidal} with $q=q_{\rm max}$, resulting in $U = 3 { \left(\frac{8\pi q_{\rm max}}{21 \text{Im}(k_2)}\right) }^{\nicefrac{1}{2}}  G^{\nicefrac{1}{4}} R_p^{\nicefrac{1}{4}} M_*^{-\nicefrac{1}{4}}  a^{\nicefrac{3}{4}}$, whereas it is simply $U = 3e_{\rm max} GM_*R_p^2 a^{\nicefrac{3}{4}} $ for the other values.}
The dashed line at 0.07 M$_\odot$ indicates the limit between red and brown dwarfs, assuming solar metallicity.
At larger semi-major axes, stronger tides are possible for eccentricities greater than 10\% but the planets would probably not be synchronously rotating.
In other words, while atmospheric tides can in principle be much stronger than shown here, a synchronously rotating terrestrial planet with a solid surface can only have atmospheric tides of 707 N~m~kg$^{-1}$, 884$\times$ greater than Earth's and 6$\times$ greater than Titan's.}
\label{fig:avm}
\end{figure}

We adopt $q_{\rm max}$ = 2.5 W/m$^2$, the value for Io, the body with the strongest tidal heating in the solar system, with intense volcanism but still a solid surface \citep{Sega:88}, and maximal eccentricity $e_{\rm max}=0.1$ \citep[a realistic upper limit for 1:1 spin-orbit resonance, i.e., synchronous rotation;][]{Riba:16},
and the imaginary component of the tidal Love number is $\text{Im}(k_2)$ = 3$\times$10$^{-3}$ \citep{Dris:15}.
Then, the maximal surface tidal potential of an Earth-sized planet with a solid surface is $U_{\rm max}$~=~707~N~m~kg$^{-1}$.
Figure~\ref{fig:avm} shows the tidal potential for an Earth-sized planet as a function of the mass of, and distance to, the primary; $U_{\rm max}$ is shown in red.
The tidal potential is mostly sensitive to the star--planet distance because of the $a^3$ in Equation~\ref{eq:tipotential}.
The maximal tidal potential intersects the habitable zone for \edit1{low-mass stars} at semi-major axes of 0.05--0.07~AU.
Equation~\ref{eq:psamplsynchro} gives a maximal change of surface pressure of 14~hPa for an Earth-like atmosphere (1~bar, N$_2$, 300~K) and an Earth-sized planet, comparable to variations at the synoptic scale of low- and high-pressure systems on Earth. 

Table~\ref{tab:comparison} compares the amplitude of atmospheric tides in the solar system and on selected M~Earths.
In all respects, the tidal wave of the Sun (or the Moon for that matter) is negligible on Earth in comparison to common meteorological phenomena.
Titan has an amplitude of 0.68~hPa, smaller than the value of 1.5~hPa found by \citet{Toka:02}, but on the same order of magnitude.
Due its larger radius, GJ~1132~b is subject to higher tides than Proxima~b, despite being farther from the $U_{\rm max}$ tidal potential in the mass-distance diagram of Figure \ref{fig:avm}.
Its surface pressure amplitude is nevertheless smaller than Proxima~b, owing to its much higher surface temperature.

\begin{table}[h!]
\nolinenumbers
\hspace{-3cm}
\begin{tabular}{p{4cm}|c|c|c|c|c|c|}
     & Earth $^{\rm 1,2}$ & Titan $^{\rm 1,2}$ & TRAPPIST-1 c $^{\rm 3,4}$ & TRAPPIST-1 d $^{\rm 3,4}$  & Proxima b $^{\rm 5}$ & GJ 1132 b $^{\rm 6,7}$ \\
            \hline
   Semi-major axis (AU) & 1 & 0.0082 & 0.0158 $^{+0.00013}_{-0.00013}$ & 0.02227 $^{+0.00019}_{-0.00019}$ & 0.0485 & 0.0153 $^{+0.0015}_{-0.0015}$\\
   Mass of the primary (M$_\Sun$) & 1 & 2.857 x 10$^{-4}$ & 0.089 & 0.089 & 0.12 & 0.181 \\
   Radius (R$_\Earth$) & 1 & 0.404 & 1.097 $^{+0.014}_{-0.012}$ & 0.788 $^{+0.011}_{-0.01}$ & 1 & 1.13 $^{+0.11}_{-0.11}$ \\
   Eccentricity & 0.0167 & 0.0292 &  0.003 $^{+0.005}_{-0.003}$ & 0.006 $^{+0.002}_{-0.002}$ & 0.1? & 0.01?\\ 
   \underline{Tidal potential (m$^{2}$~s$^{-2}$)} & 1.6 & 12.1 & 393.3 $^{+709.4}_{-393.3}$ & 144.9 $^{+108.2}_{-73.1}$ & 507.8 & 656.6  $^{+416}_{-250.3}$ \\
   \underline{Equivalent height (m)} & 0.16 & 8.9 & 37 $^{+65}_{-37}$ & 24 $^{+16}_{-12}$ & 51 & 56 $^{+54}_{-28}$  \\
   Composition  & N$_2$ & N$_2$ & CO$_2$ & N$_2$ &  N$_2$  &  CO$_2$  \\
   Surface temperature (K) & 300 & 90 & 350 & 280 & 235  &  700 \\
   \underline{Surface pressure tide (hPa)} & 0.014 & 0.68 & 9.0 $^{+16.4}_{-9.0}$ & 2.6 $^{+2.0}_{-1.3}$ & 11.0 & 7.5 $^{+2.4}_{-1.9}$ \\

\end{tabular}
\caption{Impact of Gravitational Tides on the Atmospheres of Selected Planets and Moons, assuming synchronous rotation. \edit1{Underlined parameters denote computed values.} The surface pressure tides are calculated for a  1 bar CO$_2$ atmosphere for TRAPPIST-1~c and GJ~1132~b, and a 1 bar N$_2$ atmosphere for the other bodies.
The amplitude of the surface pressure tide is very sensitive to uncertainties in orbital and planetary properties due to the exponential dependence of surface pressure on tidal potential in Equation \ref{eq:psamplsynchro}.
Note that the uncertainties of Proxima b's main parameters are not well known and therefore not considered in this table, and that the eccentricities of Proxima b and GJ 1132 b are poorly or not constrained, so reasonable upper limits are assumed.\\
{\footnotesize \edit1{References.} \edit1{(1) \citet{Arch:18}; (2) \citet{Murr:99}}; (3) \citet{Grim:18}; (4) \citet{Agol:21}; (5) \citet{Riba:16}; (6) \citet{Bert:15}; (7) \citet{Bonf:18}}}
\label{tab:comparison}
\end{table}

\section{Other configurations of gravitational tides} \label{sec:others}

As stated earlier, we focus on the simple case of an M~Earth in synchronous rotation about its star.
However, other configurations may substantially modify the gravitational tide, which we describe here.

First, planetary systems around \edit1{low-mass stars} are very compact, and planets can gravitationally influence each other, as discussed by \citet{Wrig:18} and \citet{Ling:18}.
A good example is the TRAPPIST-1 system, with seven planets orbiting their stars between 0.01 and 0.06 AU~\citep{Gill:17,Lug:17}.
The strongest atmospheric planet-planet tide in the TRAPPIST-1 system is the one raised by \mbox{TRAPPIST-1~c} upon TRAPPIST-1~b, with a tidal force 1/5\textsuperscript{th} of the Titan-Saturn one and an overpressure of $0.17 \%$. 
Therefore, planet-planet tides may also impact the atmospheric circulation.
Second, nonsynchronous spin-orbit resonance (e.g., 3:2, 2:1) may exist but is less likely to occur \citep[e.g.,][]{Murr:99}. In that case, the full tidal potential impacts the atmosphere, with the tidal potential directly proportional to $GM_*/d^3$, rather than $3eGM_*/d^3$.
Third, significant librations of the planet spin, caused by chaotic gravitational interaction in a compact system \citep{Vins:19}, may further enhance the time-dependent part of the gravitational tide.
All in all, these effects would enhance the gravitational tide, but for the sake of simplicity, we do not address them in the current study and focus only on the case that is most likely to occur and be observed.

\section{GCM simulations} \label{sec:simus}
\subsection{Setup}
The GCM used in this study is the LMD (Laboratoire de M\'et\'eorologie Dynamique) GCM \citep{Forget1999,Word:11}.
\edit1{The dynamical core solves for the hydrostatic primitive equations.
The hydrostatic approximation filters out vertically propagating sound waves and gives an unphysical behavior for gravity waves with a small horizontal wavelength \citep[][Chapter 2]{Vallisbook}.
Tidally forced internal gravity waves are unlikely to be affected, as they are expected to be long wavelength ($\sim$10$^3$~km) and not acoustic.}
The GCM is in the same configuration as in \citet{Turb:16}, who explored the possible climate of Proxima Centauri b.
Besides being the closest known exoplanet to Earth and lying in the habitable zone, Proxima Centauri b is an interesting case because it is close to the maximal tidal potential $U_{\rm max}$ (Figure~\ref{fig:avm}).
We simulate an ocean-surface M~Earth planet with an orbital eccentricity of 0.1, \edit1{where we include instellation variations due to the eccentricity}, a synchronous rotation (1:1 spin-orbit resonance), \edit1{and zero obliquity}.

\edit1{Similarly to \citet{Turb:16},} two atmospheric configurations are explored: one with an Earth-like 1 bar of N$_2$ with \edit1{an arbitrary value of} 376 ppm\edit1{v} CO$_2$ \edit1{corresponding to Earth's volume mixing ratio of CO$_2$ at the beginning of the 21st century \citep{Warn:05}}, and another with 5 bars of CO$_2$ providing an enhanced greenhouse.
\edit1{The two configurations are chosen to represent different climates, with the N$_2$ atmosphere having a frozen nightside and a habitable dayside, and the CO$_2$ atmosphere subject to a strong greenhouse effect, with a reduced day-night temperature contrast.}
\edit1{The radiative transfer of the GCM uses correlated k-distributions \citep{Fu:92,Eyme:16}, a versatile method able to simulate variable atmospheric compositions of CO$_2$, N$_2$, and H$_2$O.}
The simulations include H$_2$O liquid and ice clouds and their radiative effects but
assume a surface covered by unmoving water, i.e., with neither oceanic tides nor ocean currents.
The atmosphere exchanges water with the slab ocean by evaporation and precipitation.
Both configurations are run for 10 Earth years and result in eastward superrotating winds.
If the surface temperature drops below the freezing point, sea ice forms and grows, and is arbitrarily limited to a 1 m thickness, roughly accounting for the missing oceanic heat transport \citep{Yang:14}.
In the simulation with 1 bar of N$_2$, sea ice forms on the permanent nightside \citep[an ``eyeball'' planet;][]{Pier:10}.

We include the tide as an extra term in the equation of momentum in the GCM dynamical core:
\begin{linenomath*}
\begin{align*}
    \frac{du}{dt} & =  \left(\frac{du}{dt}\right)_{\rm dyn} - \frac{1}{R_p\cos\theta}\frac{dU}{d\lambda} \\
    \frac{dv}{dt} & =  \left(\frac{dv}{dt}\right)_{\rm dyn} - \frac{1}{R_p}\frac{dU}{d\theta},
\end{align*}
\end{linenomath*}
with $U$ the tidal potential (equation \ref{eq:tipotential}), $u$ and $v$ the zonal and meridional components of the wind, \edit1{and dyn denoting quantities calculated by the dynamical core solving for the primitive equations without gravitational tides.}
We note that the standard momentum expression is a nonlinear partial differential equation as a result of the total, or Lagrangian, derivative.
For each configuration, we run simulations with various values of tidal potential (0, $U_{\rm max}$, and 10$\times U_{\rm max}$). \edit1{With a zero potential value, we turn off the gravitational tides, so that by comparing simulations, we can separate the effects of gravitational tides from other waves, such as thermal tides that dominate the surface pressure field of planets in synchronous rotation \citep{Leco:15,Aucl:19}.}

\subsection{Surface pressure and winds}
\begin{figure}[ht]
\centering
\includegraphics[width = \textwidth]{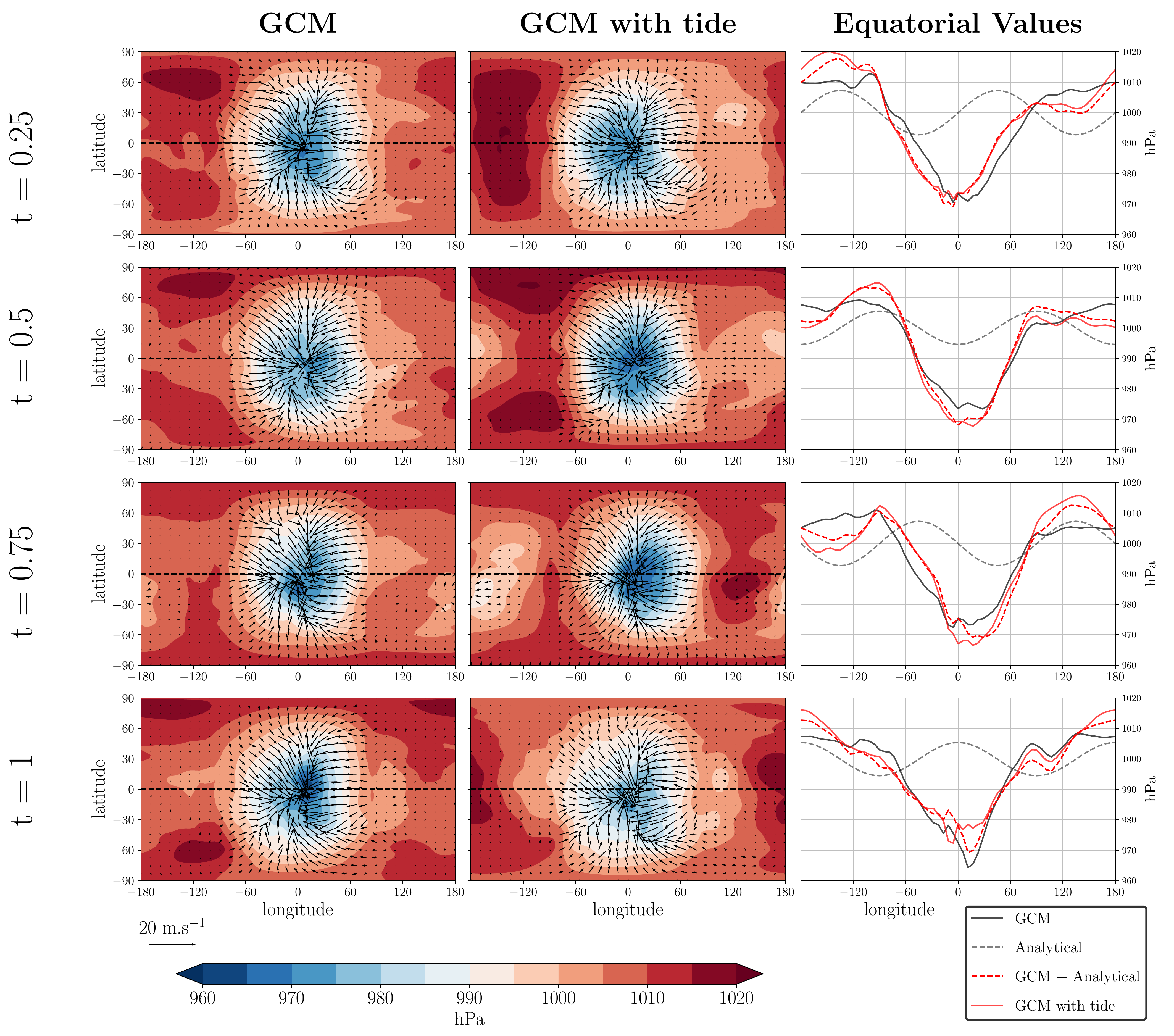}
\caption{Four snapshots of GCM surface pressure and tides.
\edit1{Instantaneous fields of} surface pressure (color contours in hPa) and surface winds (black arrows) are shown over the course of a year for a synchronous rotating planet with eccentricity $e=0.1$ without gravitational tides (left) and with tidal potential $U_{\rm max}$ (middle).
In the right panels, the equatorial values of the GCM simulations are compared to Equation~\ref{eq:ampl2}. The \edit1{variations of the} simplified analytic formulation of a gravitational tide (dashed gray line) is added to the simulated surface pressure without a gravitational tide (solid black line) to obtain a prediction (dashed red line) that is in very good agreement with the fully simulated gravitational tide (solid red line). This suggests that atmospheric tides are acting in a linear fashion.}
\label{fig:ps}
\end{figure}

The tidal effect on surface pressure and surface winds for GCM simulations with 1 bar of N$_2$ is shown in Figure~\ref{fig:ps}.
Surface winds converging at the substellar point create a low pressure at the permanent dayside. 
The gravitational tides create an eastward-moving zonal wavenumber 2 pressure wave on the fixed,  approximately wavenumber 1, low-pressure dayside system.
The gravitational tide is a residual time-dependent tide caused by the librational and radial tides.
It results in one high and one low tide per orbit. 
It is worth noting that it is unlike Earth-Moon tides, which have two highs and two lows once a period.

Figure~\ref{fig:ps} compares equatorial surface pressure for a simulation with and without tides.
The main feature is a permanent dayside low-pressure system, with winds converging toward the substellar point.
For the simulation with a tidal potential $U_{\rm max}$ (middle panels of Figure~\ref{fig:ps}), this constant surface pressure pattern is superimposed with the gravitational tide that sometimes strengthens ($t=0.5$ \edit1{-- half an orbital cycle}) and other times  weakens ($t=1$ \edit1{-- full orbital cycle}) the basic day--night pattern.
\edit1{
The effect of the gravitational tide is noticeable on surface pressure---comparable to weather systems on Earth---but is less than the effect of the thermal tide's low-pressure system on the dayside. The gravitational tide thus does not significantly affect the surface winds, which are dominated by the thermal tide.}

By adding our analytic expression of Equation~\ref{eq:psampl} (using the amplitude of Equation~\ref{eq:ampl2}) for the atmospheric tides to the GCM simulation without tides, we obtain a good match to the GCM simulation run with tides; this suggests that tides are a linear effect in this regime \edit1{with less than 1\% relative maximal amplitude of surface pressure variations}, despite the fact that they modify the nonlinear momentum equation.

Surface winds are not significantly impacted by the gravitational tide. The day-night surface temperature gradient is a robust feature of synchronously rotating M~Earth, and remains the driving mechanism for surface winds generally oriented toward the substellar point (Figure \ref{fig:ps}). This result differs from Titan, where \citet{Toka:02} found surface tidal winds, owing to the low insolation and temperature contrast at the surface of Titan.

\section{Impact of tides on the simulated Climate}\label{sec:results}

\subsection{Atmospheric temperature response}

\begin{figure}[ht]
\centering
\includegraphics[width = \textwidth]{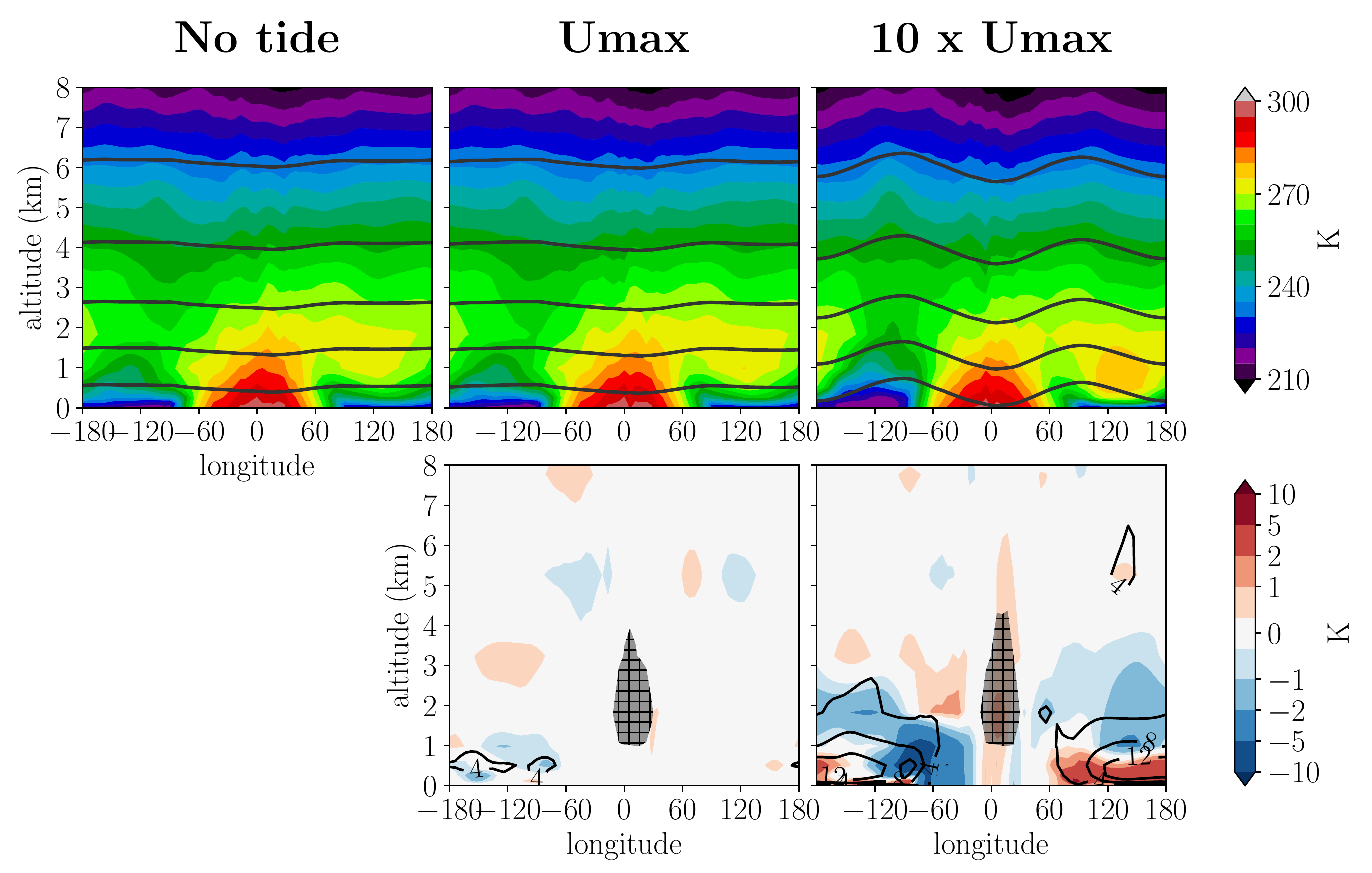}
\caption{
Equatorial atmospheric temperature structure.
Temperature (color contours) and isobars (black contours), averaged for 5$^\circ$S--5$^\circ$N, are shown as a function of longitude and altitude for a simulated Proxima b with a 1 bar, N$_{\rm 2}$--dominated atmosphere.
Simulations without a gravitational tide (left), with a $U_{\rm max}$ (middle), and with a $10 \times U_{\rm max}$ (right) gravitational tide are shown.
The bottom plots show the average (color contours) and standard deviation (black contours) of the temperature difference between the case with and without tides over four years.
The grey shaded area shows the location of H$_2$O clouds ($\ge$ 20 ppm).
The average atmospheric temperature caused by the $U_{\rm max}$ tide rarely exceeds 1~K, whereas the $10 \times U_{\rm max}$ tide causes a change of temperature up to 10~K, in particular near the surface on the nightside.
These changes in temperature do not affect the formation of dayside clouds, and thus do not create any radiative cloud feedback.
The Weak Temperature Gradient (WTG) theory still holds for the $U_{\rm max}$ tide but starts to break down for a $10 \times U_{\rm max}$ tide.
}
\label{fig:warming}
\end{figure}

From the horizontal momentum equation, the Weak Temperature Gradient theory \citep[WTG;][]{Pier:19} states that a strong gradient in insolation (or any thermal forcing) results in a strong circulation but weak temperature and pressure gradients.
Applied to synchronously rotating planets, the temperature structure above the planetary boundary layer is homogenized \citep[e.g.,][]{Merl:10}.
Figure \ref{fig:warming} shows the atmospheric temperature for the N$_2$ atmosphere without any tidal effect, with a global variation of 20~K at 2~km, in contrast to the 150~K difference between the dayside and nightside surface temperatures.
The WTG theory still holds true with tidal potential $U_{\rm max}$.
However, the WTG breaks down in a hypothetical case with $10\times U_{\rm max}$, with heat transported eastwards by the tide and temperature levels following perturbed pressure levels.

The simulation with a $10\times U_{\rm max}$ tidal potential is an order of magnitude beyond the ``maximal'' tides for an M~Earth. This stronger tidal case is worth discussing for two reasons: (1) a planet with a greater tidal $Q$ could, in principle, have relatively low internal heating for a given tidal forcing and, (2) we are interested in the atmospheric response to strong tides, even for nonterrestrial planets.  While we save a complete study of strong atmospheric tides for future studies, we present here a first look at that regime. With $10\times U_{\rm max}$, the tide transports even more heat eastwards.
This transport can be clearly seen in Figure~\ref{fig:warming}, with more than 5~K of cooling and warming, westwards and eastwards respectively.
At an altitude of 250~m, the temperature exceeds 273~K on the east side of the day-night terminator (longitudes 100E to 180E) for the case with $10\times U_{\rm max}$, or is below 230~K on the west side. 
However, even with a tide multiplied by 10, the impact on sea ice area is negligible, with surface temperatures below the freezing point at all locations on the nightside after 150 orbits of simulation.

\begin{figure}[ht]
\centering
\includegraphics[width = \textwidth]{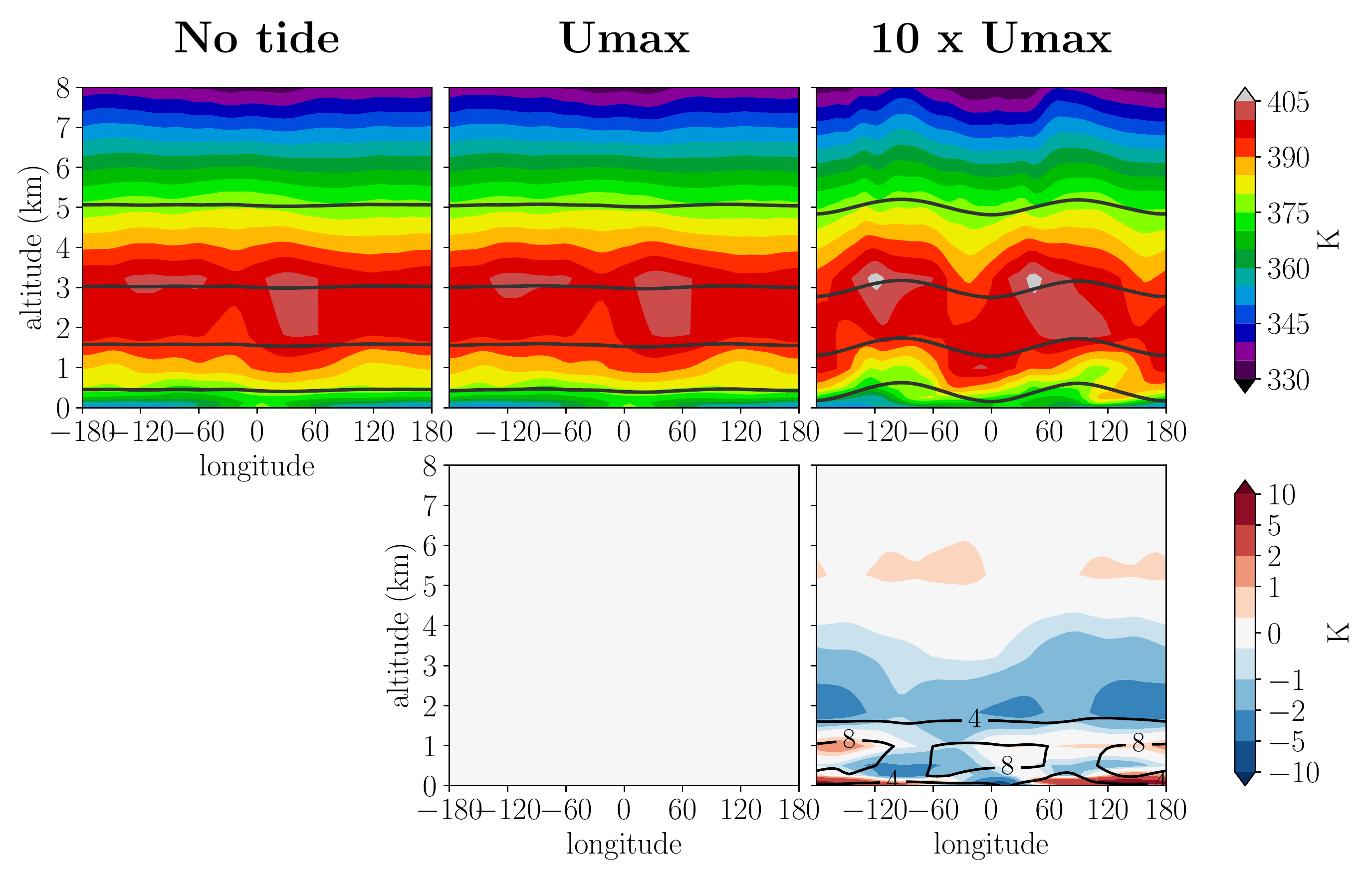}
\caption{Same as Figure \ref{fig:warming} but for a 5-bar CO$_{\rm 2}$--dominated atmosphere.
There is less than 0.5~K difference for the realistic $U_{\rm max}$ tide, whereas the $10 \times U_{\rm max}$ tide causes a cooling of more than 2~K centered at a 2~km altitude.
Note that clouds form at a higher altitude of 10 km, and are not impacted by tides, as in the case of a 1 bar N$_{\rm 2}$ atmosphere.}
\label{fig:warmingCO2}
\end{figure}

Figure \ref{fig:warmingCO2} shows the same behavior for another simulation with a 5~bar CO$_2$ atmosphere. Regardless of a gravitational tide, the thermal gradient is weak near the surface, as all surface ice is melted and surface temperatures are well above the freezing point everywhere on the planet. Owing to the strong greenhouse effect and heat transport by the eastward superrotating atmosphere, the temperature structure does not depend on substellar longitude in the troposphere with or without a $U_{\rm max}$ tide.
Interestingly, the artificially increased $10 \times U_{\rm max}$ tide creates a dependence on atmospheric temperatures with longitude, but also causes a net cooling in the lower atmosphere.
At 2 km of altitude, the average temperature is 2~K lower because of the tide.
A possible mechanism causing this cooling is the nonlinearity of the Stefan-Boltzmann law \citep{Henr:19}, as local tidal variations of pressure along the orbit create adiabatic fluctuations in the temperature, resulting in a net increase of radiated power.
This, in turn, cools the atmosphere, with this effect being more pronounced for the warmest atmospheric layers, as per the $\sigma T^4$ Stefan-Boltzmann law.
The reason for the location and magnitude of the temperature changes induced by the gravitational tide and Stefan-Boltzmann law is a combination of the vertical radiative balance and convective influences on the lapse rate \citep[e.g.,][]{Koll:16}.

\begin{figure}[ht]
\centering
\includegraphics[width = \textwidth]{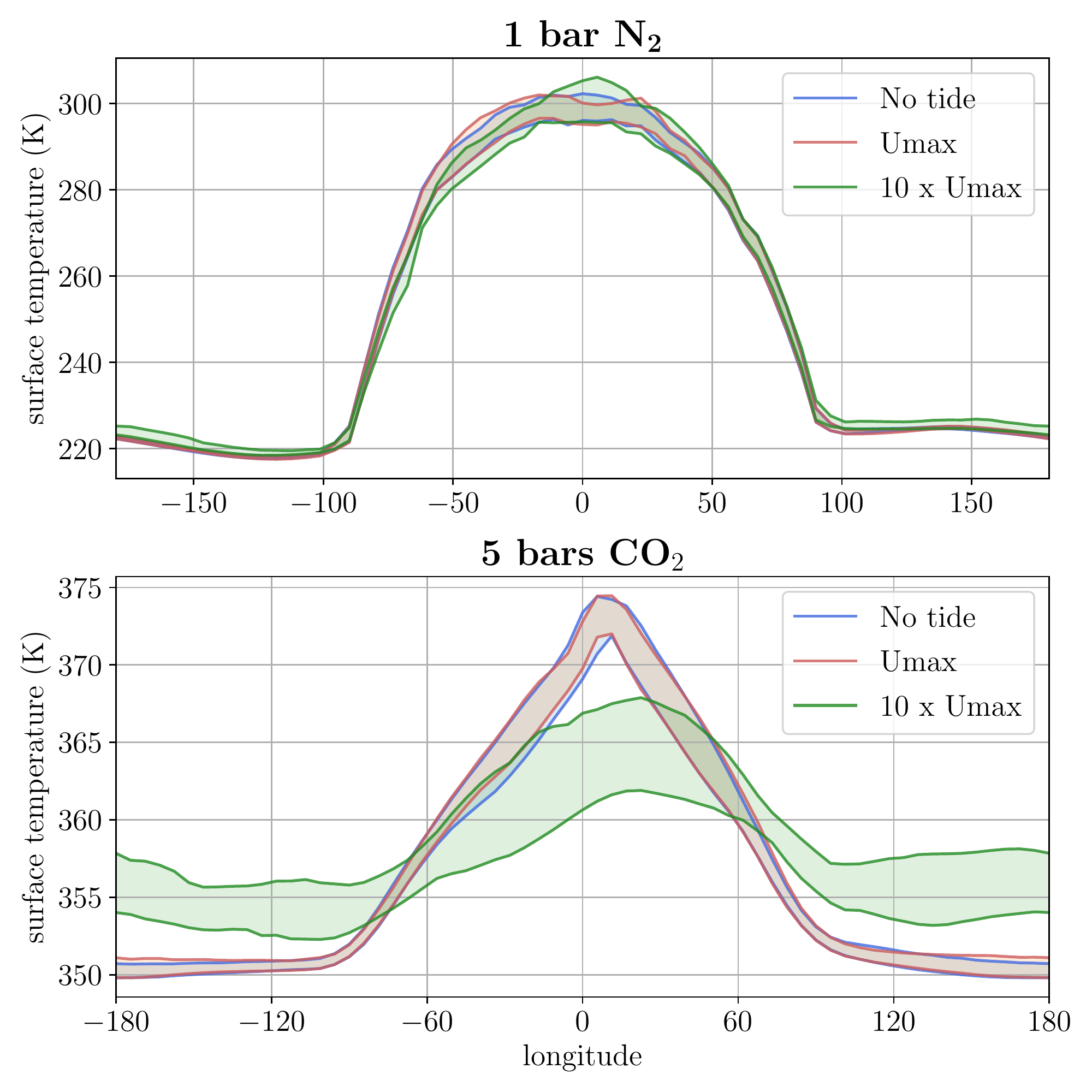}
\caption{
Equatorial surface temperature.
The surface temperature range over two orbits, averaged for 5$^\circ$S--5$^\circ$N, is shown as a function of longitude for a simulated Proxima b with a 1 bar, N$_{\rm 2}$--dominated atmosphere and a 5 bar, CO$_{\rm 2}$-- dominated atmosphere.
The surface temperature change caused by the $U_{\rm max}$ tide is negligible in comparison to the variability of surface temperature without a tide.
The $10 \times U_{\rm max}$ tide causes a wider range of surface temperature variations.
In the case of a 5 bars CO$_{\rm 2}$-- dominated atmosphere, the average surface temperature contrast between day and night is lowered, due to the $10 \times U_{\rm max}$ tide's transport of heat.
For a 1 bar, N$_{\rm 2}$-- dominated atmosphere, the day-night contrast is not impacted by the tide because the heat transport is not as strong.
}
\label{fig:surft}
\end{figure}

\edit1{The surface equatorial temperature is also impacted by the heat transport, with reduced day-night contrast. This is especially apparent in figure~\ref{fig:surft} for the 5 bar CO$_2$ atmosphere, with the average difference decreasing from 23~K without a tide, to 10~K with a  $10 \times U_{\rm max}$ tide. This case is the one with the most heat transport, compared to the other cases that show negligible impact. The gravitational tide highs and lows also increase the fluctuation range of surface temperature at a given location, as shown in figure~\ref{fig:surft}.}

\subsection{Clouds}

Next, we explore the possibility of a cloud feedback mechanism triggered by gravitational tides, which may in turn affect the climate.
It could be naturally expected that the ebb and flow of pressure create or dissipate clouds, hence impacting the climate.
However, we find that in all cases, even $10\times U_{\rm max}$, a gravitational tide does not impact H$_2$O cloud formation or abundance in a way to appreciably affect the mean climate.
In fact, and despite the tide-induced pressure variations, cloud formation is rarely affected by the tides. Using the hydrostatic equilibrium
\( \displaystyle \frac{dp}{p} = -\frac{mg}{RT}dz \)
with Equation~\ref{eq:dpsurp}, we obtain the equivalent height of the horizontal pressure bulge
\( \displaystyle z_{eq} = Ug^{-1}\).
Air flowing over the topography of height $z_{eq}$ gives a sense of the variations of local atmospheric conditions due to tides.
This height remains modest, as given in Table \ref{tab:comparison}, with a wave amplitude of a few tens of meters over a year.
For instance, for a planet with Earth's size and mass, we get from $U_{\rm max}$ that $z_{eq} \leqslant$~72~m.

The two types of atmospheres explored in this article provide contrasting examples of potential cloud behavior.
In the N$_2$--dominated atmosphere, H$_2$O clouds are principally formed by moist convection within 4 km of the surface on the dayside. Low-altitude clouds primarily affect shortwave radiation because they re-emit longwave from temperatures close to that of the underlying surface.
This strong pattern of cloud formation remains present regardless of the gravitational tide.
The tide creates a change of less than 1~K for the $U_{\rm max}$ case, negligible for cloud formation.
The $\sim$5~K cooling westwards, and warming eastwards, of the permanent dayside caused by the $10\times U_{\rm max}$ case do not occur where dayside cloud forms, and thus do not create any cloud feedback. 
Any cloud radiative feedback in the shortwave remains negligible.

In the CO$_2$--dominated atmosphere, clouds form above the 10 km height on both day and night sides, but the impact of the gravitational tide on the planet's climate is also negligible, with a less than 0.5~K difference on average.
In the infrared, cloud radiative effects depend on cloud altitude and strengthen atmospheric greenhouse as they absorb and re-emit at cold temperatures. 
Tides are stronger at lower altitudes because they depend on atmospheric mass, so clouds remain unaffected in this configuration.

We do not exclude that cloud feedback driven by gravitational tides may exist, either for colder surface temperature M~Earths in the shortwave by triggering condensation of near-surface clouds, or in the infrared by changing cloud altitude.
However, we did not find either configuration while we explored various surface pressures (0.1 to 5 bars), compositions (N$_2$ and CO$_2$), and tidal potentials (from 0 to $10\times U_{\rm max}$).

\section{Conclusions} \label{sec:conclusion}

We have assessed the atmospheric effects of gravitational tides for tidally locked, synchronously rotating terrestrial exoplanets \edit1{with a static ocean} in eccentric orbits around M dwarfs. These tides are orders of magnitudes stronger than for any planetary or lunar atmosphere in the solar system. 
We have found an analytical solution for surface pressure changes, and an upper limit for gravitational tides of M~Earths with a solid surface.
We then implemented tidal effects in a state-of-the-art exoplanet GCM for the first time, and validated the analytical formulation against GCM results.
Gravitational tides create pressure waves, with associated temperature changes that affect the local meteorological conditions over an annual period (e.g., surface pressure high and lows, slightly enhanced precipitation, and changes in surface temperature).
GCM simulations revealed two possible modest impacts of gravitational tides on the climate: longitudinal transport of heat with the tide, and cooling of the hot lowest atmospheric layers by the enhanced emission of radiation.
In our simulations, the gravitational tides had a limited effect on both low- and high-altitude clouds.

\edit1{While not simulated in this study, we can postulate that in the case of a dynamic ocean, more heat is transported by the oceanic circulation to the nightside \citep{Yang:14}, further reducing the day-night contrast in addition to gravitational tides. In the case of land, mountain ranges would limit the transport of heat and water by rains on the windward side, possibly decreasing the transport of heat by gravitational tides. As explained in section \ref{sec:potential}, we have ignored that the deformation of the solid body changes the tidal potential exerted on the atmosphere. As oceans are more deformable by gravitational tides than land, the more a planet is covered by oceans, the more the tidal potential of Equation \ref{eq:tipotential} is reduced, and the smaller the surface pressure amplitude of Equation \ref{eq:psamplsynchro}.}

These results are part of the community's long-term efforts to characterize conditions at the surface of M~Earths, planets with no equivalent in the solar system.
The lack of observational constraints for M~Earths makes them subject to speculation on the characteristics and diversity of their climate and surface conditions. 
So far, their climate and meteorology are explored through theoretical and numerical studies,
revealing the complex impact of a permanent nightside and dayside on atmospheric circulation and water trapping \citep[e.g.,][]{Meno:13,Yang:14}.
Among the countless phenomena that may affect such climates, gravitational tides are one topic that had not been tackled until now, as we have developed a modeling solution to explore those effects.
In the solar system, Titan has an atmosphere subject to the highest gravitational tides, with a probable impact on the circulation as inferred from theory \citep{Lore:92}, general circulation models \citep{Toka:02,Charn:12}, and observations \citep{Rodr:11}.
Therefore, Titan's atmosphere could appear as a starting point for assessing atmospheric gravitational tides.
However, we have shown that gravitational tides on M~Earths only moderately impact their surface meteorology, with little to no impact on their climate, despite much stronger tides than on Titan.
Our results suggest that a key difference between M~Earths and Titan is the strong radiative forcing that drives day-to-night circulation.
Solar radiation on Titan has a negligible impact on its circulation compared to the gravitational tides of Saturn.
In contrast, the strong differences in day and night near-surface temperatures and pressures of synchronously rotating M~Earths exceed the impact of gravitational tides.

\edit1{The tide-locking time of close-in planets can range from less than ten to hundreds of millions of years \citep{Pier:19}. During the capture into a final spin-orbit state, gravitational tides may impact the climate, e.g., heat transport toward the nightside, and in turn, the geophysical evolution of the planet. A full picture of the possible impacts of gravitational atmospheric tides on a planet's history is another largely unexplored topic. This would require first exploring such tides for spin configurations other than 1:1.}

A possible effect of gravitational tides, not investigated here, is the mesoscale (typically $\le$ 1000 km) interactions with topography.
On Earth, tides create currents at the bottom of the oceans, where their interaction with the ocean-surface topography causes vertical displacements, a source of waves propagating in the oceans \citep{Garr:07}.
As deep oceanic tides enhance mixing and dissipate tidal energy, similar atmospheric effects may be significant for atmospheric dynamics of planets with high gravitational tides and surface topography.

\section*{Acknowledgments}
We thank two anonymous reviewers for their helpful suggestions. We are grateful to Martin Turbet for providing converged atmospheric states of Proxima Centauri b. T.N. is supported by the McGill Space Institute. We acknowledge a Compute Canada/Canada Foundation for Innovation computing allocation.

\bibliography{AtmoTide}{}
\bibliographystyle{aasjournal}

\end{document}